\documentclass[aps,pre,twocolumn,amsmath,amssymb,nofootinbib,longbibliography,superscriptaddress]{revtex4-1}
\pdfoutput=1
\usepackage[latin9]{inputenc}
\usepackage{braket}
\usepackage{xcolor}
\usepackage{verbatim}
\usepackage{graphicx}
\usepackage{esint}
\usepackage{epstopdf}
\usepackage{float}
\usepackage{bm}
\definecolor{darkblue}{rgb}{0,0,0.6}
\usepackage[colorlinks,linkcolor=darkblue,citecolor=darkblue,urlcolor=darkblue]{hyperref}

\newcommand{\beq}{\begin{equation}}
\newcommand{\eeq}{\end{equation}}

\let \a=\alpha

\begin{document}

\title{Yielding and plasticity in amorphous solids}

\author{Ludovic Berthier}

\affiliation{Laboratoire Charles Coulomb (L2C), Universit\'e de Montpellier, CNRS, 34095 Montpellier, France}

\affiliation{Gulliver, UMR CNRS 7083, ESPCI Paris, PSL Research University, 75005 Paris, France}

\author{Giulio Biroli}

 \affiliation{Laboratoire de Physique de l'Ecole Normale Sup\'erieure, ENS, Universit\'e PSL, CNRS, Sorbonne Universit\'e, Universit\'e Paris Cit\'e, F-75005 Paris, France}

\author{Lisa Manning}

\affiliation{Department of Physics, Syracuse University, Syracuse, New York 13244, USA}

\author{Francesco Zamponi}

\affiliation{Dipartimento di Fisica, Sapienza Universit\`a di Roma, Piazzale Aldo Moro 5, 00185 Rome, Italy}

 \affiliation{Laboratoire de Physique de l'Ecole Normale Sup\'erieure, ENS, Universit\'e PSL, CNRS, Sorbonne Universit\'e, Universit\'e Paris Cit\'e, F-75005 Paris, France}

\date{\today}

\begin{abstract}
The physics of disordered media, from metallic glasses to colloidal suspensions, granular matter and biological tissues, offers difficult challenges because it often occurs far from equilibrium, in materials lacking symmetries and evolving through complex energy landscapes. Here, we review recent theoretical efforts to provide microscopic insights into the mechanical properties of amorphous media using approaches from statistical mechanics as unifying frameworks. We cover both the initial regime corresponding to small deformations, and the yielding transition marking a change between elastic response and plastic flow. We discuss the specific features arising for systems evolving near a jamming transition, and extend our discussion to recent studies of the rheology of dense biological and active materials. 
\end{abstract}

\maketitle

\section{Introduction}

\label{sec:introduction}

We present an overview of recent theoretical progress on the rheological response of disordered media to deformation. Disordered media, which include a wide range of materials from soft and biological matter to hard glasses, are characterized by complex and diverse mechanical behaviors~\cite{rodney2011modeling,berthier2011theoretical}. Understanding the rheology and yielding of these materials is crucial for a variety of applications~\cite{bonn2017yield}, including the design of materials with specific mechanical properties~\cite{keim2019memory} and the prediction of the response and possible failure of systems to external stimuli.

The diversity of amorphous materials can be appreciated by the range of timescales and lengthscales spanned by their elementary constituents~\cite{rodney2011modeling,berthier2011theoretical}. Metallic and molecular glasses are composed of atoms and molecules with a typical scale of a few Angstroms, moving over picosecond timescales~\cite{ediger1996supercooled}. Colloidal glasses span a range from tens of nanometers to microns and timescales from microseconds to seconds~\cite{hunter2012physics}. Granular particles and foams are composed of non-Brownian particles of macroscopic sizes, with no intrinsic dynamic timescales other than those due to the driving forces~\cite{jaeger1996granular}. Active and living matter are often made of objects of colloidal sizes moving over timescales essentially controlled by internal sources of motility that are not of thermal origin, but result from energy consumption at the local scale~\cite{marchetti2013hydrodynamics}. 

Here, we present a unifying perspective on their rheological behaviors based on recent numerical and analytical works. In fact, despite the broad range of scales, specific interactions, and microscopic dynamics listed above, an important body of works have recently shown that the rheological response of amorphous materials, especially in the limit of small rates of deformation, is quite universal. We shall show that, once appropriate rescaled units are introduced, remarkable connections between the rheological behaviors of microscopically very different amorphous materials emerge.  

\begin{figure*}[t]
\includegraphics[width=\textwidth]{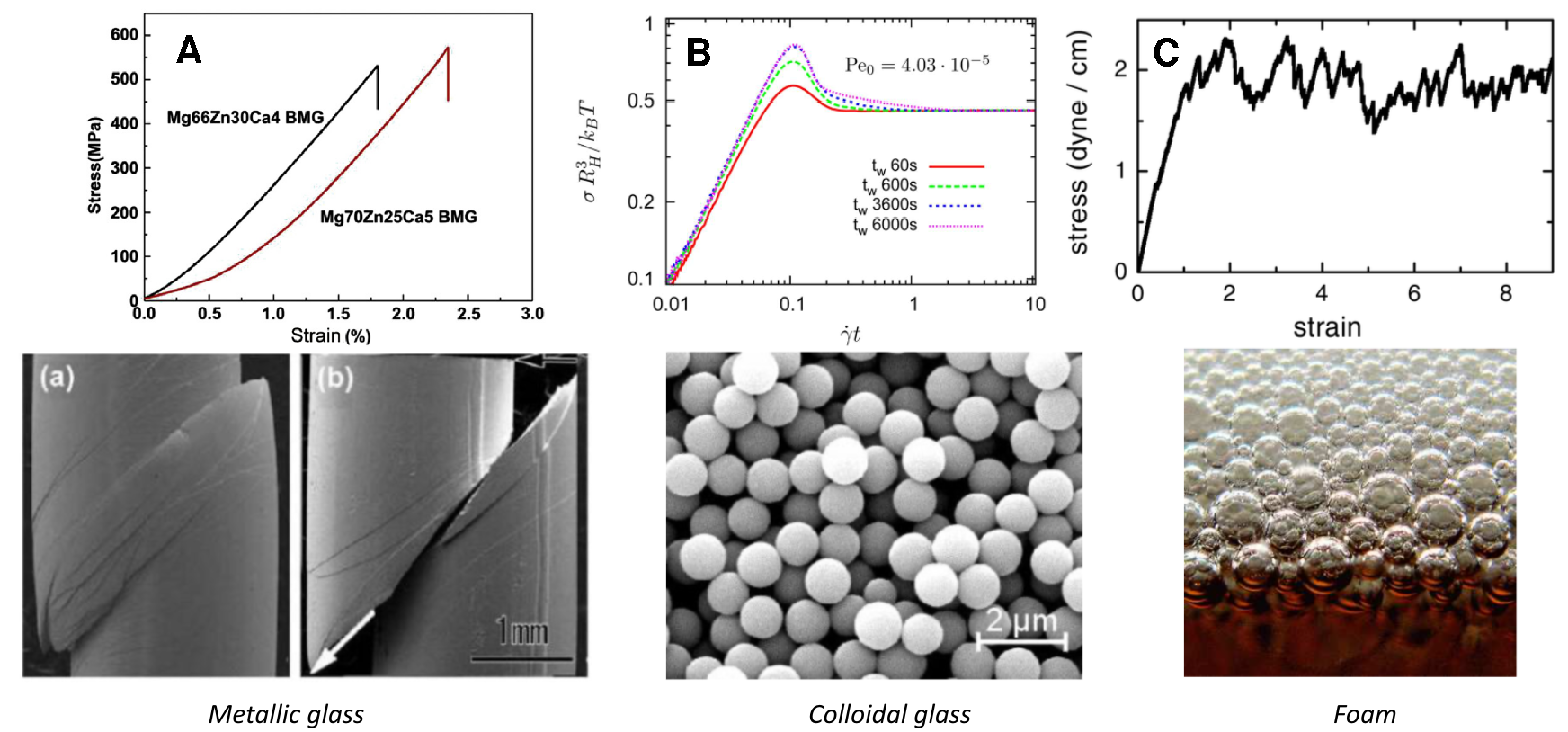}
\caption{{\bf Deformation and yielding of amorphous solids.} Selected examples of mechanical loading experiments in amorphous materials spanning a broad range of time scales, length scales, and physical behaviours. (A) Compressive test in two metallic glasses produces elastic deformation followed by macroscopic failure, from Ref.~\cite{li2016recent}. (B) Sheared colloidal suspensions display elastic response followed by plastic flow after yielding marked by a stress maximum, from Ref.~\cite{amann2013overshoots}. (C) Elastic and plastic flow in a foam, from Ref.~\cite{lauridsen2002shear}. 
}
\label{fig:intro}
\end{figure*}

A hint of this universality is illustrated in Fig.~\ref{fig:intro}, which shows a selection of mechanical tests performed on a metallic glass, a colloidal suspension, and a foam. While the geometry of the tests may vary from one material to the other (for instance compression versus simple shear), all materials respond essentially as an elastic solid at small deformation with a linear relation between stress and strain. Deviations from linearity are observed at larger deformation, suggesting that some form of plasticity occurs even in this seemingly elastic regime. The material finally yields at larger deformation amplitude, in a manner that we will carefully discuss. Past yielding, plastic flow is observed, which can sometimes lead to shear-banding in ductile soft matter systems~\cite{bonn2017yield}, or to a more abrupt failure in metallic glasses characterised by shear localisation and brittle failure~\cite{schuh2007mechanical}. 

We will 
focus on yielding, and more generally the mechanical response of disordered media under slow driving, 
and on systems that transition from a flowing to an amorphous solid phase in very different ways. Atomic, molecular and colloidal glasses undergo a glass transition as their density is increased or their temperature decreased~\cite{berthier2011theoretical}. On the other hand, non-Brownian systems acquire rigidity by crossing a jamming transition controlled by the density in the absence of thermal fluctuations~\cite{liu2010jamming}. Glass and jamming transitions are distinct phenomena, characterised and identified by different tools~\cite{mari2009jamming,berthier2009glass}. In particular, the criticality of the jamming transition has specific consequences on the rheology of jammed materials that we discuss below in a separate section. Yet, jammed and glassy materials may also share common rheological behaviour in some cases~\cite{ikeda2012unified}. It was recently recognised that dense active and biological materials can also undergo nonequilibrium glass transitions when the amplitude of the active forces is varied~\cite{berthier2019glassy}. Finally, theoretical models describing biological tissues may also display a form of jamming transition in the absence of driving forces~\cite{bi2016motility}, which endow them with remarkable rheological properties.  

The main idea underpinning the universality of rheological behaviors of amorphous materials is the concept of glass stability. This is directly related to how amorphous solids are formed, and can be used to organize a myriad of disparate results about their rheological behavior. Thinking about the complex energy landscape accessible to disordered states of matter, one realises that molecular systems occupy very deep regions of the landscape. These highly stable regions are accessible because molecular liquids transition to glasses when their collective relaxation time is about 14 orders of magnitude longer than their microscopic  relaxation time~\cite{ediger1996supercooled}, which allows them to access low-lying energy minima. Recently, molecular glasses prepared using physical vapor deposition have been shown to be located even deeper in the energy landscape, and for this reason they form ultrastable glasses~\cite{ediger2017perspective}. By comparison, colloidal glasses are prepared over timescales that are about 5 orders of magnitude slower than their intrinsic microscopic timescale~\cite{hunter2012physics}, and as a result occupy much less stable glassy states. Finally, non-Brownian systems can be thought of as occupying the highest levels of the glassy landscape as the absence of thermal fluctuations prevents its exploration~\cite{nishikawa2022relaxation}. These systems are thus even less stable. 

Historically, it has been difficult to simulate this entire range of responses on a computer, as the very deep minima accessible to molecular glasses would require simulations that run 15 orders of magnitude longer than the microscopic time scale that dictates the molecular dynamics~\cite{berthier2023modern}. However, novel Monte Carlo algorithms were recently developed to efficiently explore these different preparation protocols in computer simulations of simple models for glassy systems, without changing the type of particle interactions~\cite{berthier2016equilibrium,swap}. These developments thus provide a way to numerically study systems comparable to metallic glasses, colloids or emulsion droplets in a unified manner and compare in particular their rheological behaviours. These wildly different preparation protocols generate different types of mechanical responses and failures, even when particle interactions are the same~\cite{ozawa2018random}.  

We will first focus on the initial deformation regime leading to yielding. This regime is characterized by the elastic response of the material, as well as the approach to global failure, and is termed the pre-yielding regime. Understanding the mechanisms underlying these processes is critical for predicting the behaviour of materials under applied stress, including periodic deformation protocols.

Our second topic is the yielding transition itself, which refers to the point at which a material undergoes a significant change in behavior, roughly transitioning from elastic response to plastic flow. The yielding transition has attracted significant attention in recent years due to its fundamental importance and practical implications. We will discuss the various approaches that have been used to study this transition, including experimental and theoretical techniques.

A third area of focus will be the rheological properties of materials near the jamming transition, which is a critical point that separates the fluid-like and solid-like behavior of a non-Brownian system. We will discuss the various techniques that have been used to study the rheological properties of materials near the jamming transition, as well as the key insights that have been gained from these studies.

We then provide a perspective on various attempts to define, detect and characterize the statistical properties of localized regions in amorphous structures that eventually act as localized plasticity defects. This topic is still under intense scrutiny, boosted in particular by recent developments in machine learning techniques.   

Finally, we will discuss the relatively new field of studying the rheological properties of amorphous active and living matter. These materials, which encompass systems from active colloids to biological tissues, exhibit complex behavior that is influenced by both their internal dynamics and their interaction with the environment. Understanding the rheological properties of these materials is important for a variety of applications, including the design of materials with specific mechanical properties and the prediction of the response of biological systems to external stimuli.

Throughout this review, we emphasize the importance of a unified approach to studying the rheology and yielding of this broad range of disordered media. Such an approach involves the use of a wide range of techniques, including numerical simulations, experiments, and theoretical approaches deeply rooted in statistical mechanics. By combining them, a deeper understanding of the fundamental mechanisms underlying the behavior of disordered materials can be gained to develop more accurate predictive models. We concentrate on rheological responses in the limit of slow driving, where the behavior becomes independent of the frequency or rate of applied deformation. Other reviews have focused on how material response changes as a function of strain rate and frequency~\cite{nicolas2018deformation, voigtmann2014nonlinear}. Here we merely speculate on how emerging insight in the limit of slow driving might impact research into finite-rate behavior. By providing a broad overview of the current state of the field and highlighting key open questions, we wish to provide a useful bibliographical resource for researchers working in this area and to stimulate further research and discussions. 


\section{Before yielding: irreversibility, avalanches, memory}

In this section, we focus on the first initial deformation regime (pre-yielding), which is common to all amorphous materials. From the macroscopic point of view, the solid appears to respond almost linearly, i.e. elastically, see Fig.~\ref{fig:intro}. Microscopic studies reveal a much more complex and interesting situation, in which irreversibility, avalanches, hysteresis and memory effects play an important role.

\subsection{Irreversibility}
\label{sec:reversibility}

\begin{figure*}[t]
\includegraphics[width=\textwidth]{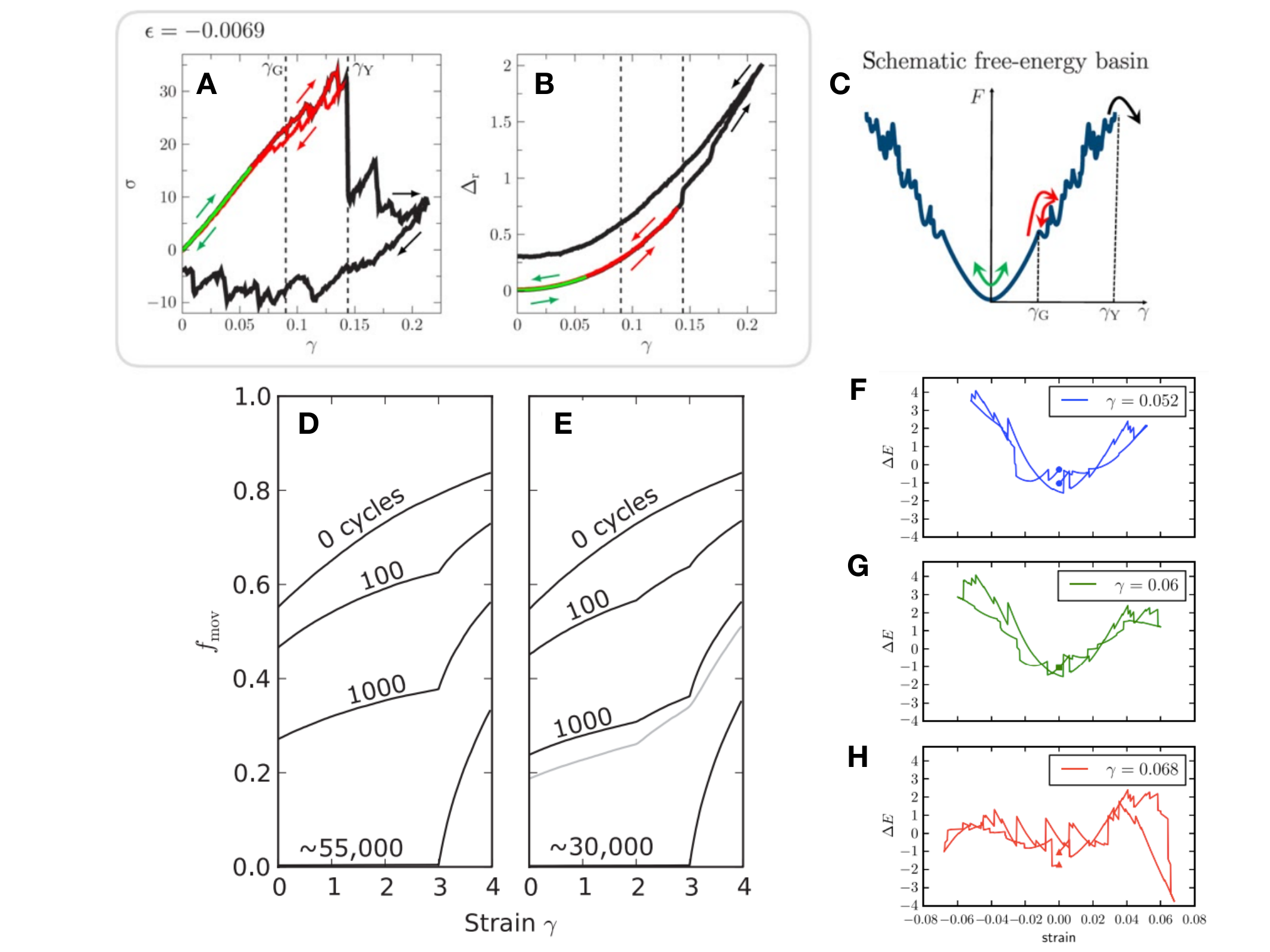}
\caption{{\bf Oscillatory strain and reversibility.}
(A) Stress $\sigma$ versus applied strain $\gamma$ in a colloidal glass up to some maximum value $\gamma_{\rm max}$, at which the strain is reversed back to zero, and (B) non-affine mean-squared displacement $\Delta_r$ between the configurations at $\gamma=0$ and at $\gamma$, both from Ref.~\cite{jin2018stability}. 
(C) Sketch of the free energy landscape, from Ref.~\cite{jin2018stability}. For small $\gamma_{\rm max}$ (green), a reversible regime has a linear stress-strain curve, and the system comes back to the original configuration ($\Delta_r=0$) when $\gamma$ is reverted back to zero.  For larger $\gamma_{\rm max}$ (red), in a partially irreversible regime, stress drops are observed, but the system still comes back to its original state at $\gamma=0$. For even larger $\gamma_{\rm max}$ (black), an irreversible yielding happens and the system escapes its original basin, hence $\Delta_r>0$ at $\gamma=0$. (D) Training a material to encode memories via cycles of fixed amplitude $\gamma_{\rm train}=3$, from Ref.~\cite{keim2011generic}. During read-out, a kink is observed at $\gamma_{\rm read} = \gamma_{\rm train}$. For a large number of training cycles, no motion is present for $\gamma_{\rm read} < \gamma_{\rm train}$, indicating full reversibility. (E) Training via alternating cycles of $\gamma^1_{\rm train}=3$ and $\gamma^2_{\rm train}=2$, from Ref.~\cite{keim2011generic}. During training, the system keeps memory of both training strains, but after many cycles, the memory of the smallest training strain is erased because perfect reversibility is observed up to the largest one. Introducing some noise (light gray curve) helps preserve both memories indefinitely. (F-H) Similar training results for models with long-range interactions, from Ref.~\cite{fiocco2014encoding}. In (G), training is performed at $\gamma_{\rm train}=0.06$ and the trajectory of the inherent structure energy is perfectly reversible during a read-out phase with $\gamma_{\rm read} = \gamma_{\rm train}$. On the contrary, if $\gamma_{\rm read} \neq \gamma_{\rm train}$ (F,H), the system does not come back to the initial state after a single read out-cycle.}
\label{fig:oscillatory}
\end{figure*}

The pre-yielding rheological behavior of amorphous solids is astonishingly complex. In this region (see Fig.~\ref{fig:intro}), the stress $\sigma$ grows, on average, almost linearly with the strain $\gamma$, so the response is on average that of an elastic solid. Moreover, if the applied strain is reversed back to zero, the system returns to its initial pre-strain configuration.
Both observations naively suggest a relatively simple solid-like elastic response at small enough applied strain (see Fig.~\ref{fig:oscillatory}A). Yet, on a mesoscopic scale (or in single samples in the case of numerical simulations), an intermittent plastic response, punctuated by irreversible stress drops, is observed.

Let us consider cyclic shear experiments and numerical simulations, in which an oscillatory strain is applied to the sample, with some given amplitude and very low-frequency~\cite{pine2005chaos,corte2008random,fiocco2013oscillatory,regev2013onset,regev2015reversibility,kawasaki2016macroscopic,jin2018stability,bhaumik2019role,yeh2020glass}. The case of vanishing frequency (quasi-static oscillatory strain) is particularly interesting. In this setting, three distinct regimes have been observed (Fig.~\ref{fig:oscillatory}A): (i) a fully reversible, elastic regime, at small amplitude; (ii) a partially irreversible regime at intermediate amplitudes, in which the system displays plastic response, but still manages to exactly revert back to its original state when the strain is removed; (iii) an irreversible regime when yielding is achieved at even larger amplitude.

The very existence of the first, reversible regime in the thermodynamic limit is questionable~\cite{hentschel2011athermal,lerner2018protocol}. In the picture where plasticity is mediated by localized defects that have a finite concentration, the probability to excite one of them by an infinitesimal applied strain goes to one when the system size increases. Yet,
in ultrastable glasses, the minimal value of $\gamma$ at which the first plastic event is observed decreases very slowly with $N$~\cite{lerner2018protocol}, and as a consequence the reversible regime seems to persist up to extremely large system sizes. The yielding transition itself will be discussed in Sec.~\ref{sec:yielding} and we now focus on the partially irreversible regime.

\subsection{Avalanches}

The existence of a partially irreversible regime is intimately related to the presence of plasticity defects, and its characterization is theoretically challenging. Due to the structural disorder of the glass configuration, it is very hard to distinguish a plastic defect from an equally disordered background; we will discuss this problem in more details in Sec.~\ref{sec:defects}, together with the microscopic nature, structure and density of these defects. Moreover, while the non-affine displacement that is observed at plastic events is initially localized around defects, the relaxation of individual defects may itself trigger, via elastic interactions, the creation of new defects, whose relaxation may lead to a large-scale avalanche. In this section, we review recent efforts devoted to the characterization of plasticity in the pre-yielding regime, and in particular the avalanches, their statistics, and the associated critical exponents~\cite{maloney2004subextensive,karmakar2010statistical,talamali2011avalanches,hentschel2011athermal,antonaglia2014bulk,lin2014scaling,lin2014density,lin2016mean,franz2017mean,lagogianni2018plastic,lerner2018protocol,ruscher2020residual,shang2020elastic,franz2021surfing}.

We begin by briefly reviewing the scaling arguments developed in Ref.~\cite{lin2014scaling,lin2014density,lin2016mean}, see also Ref.~\cite{shang2020elastic}. Suppose that the stress-strain curve is composed of piecewise linear 
regions where $\Delta\sigma \sim \mu \Delta\gamma$ ($\mu$ is the elastic modulus defined in the truly linear regime), separated by stress drops due to plastic avalanches. Next, suppose that at a given value of the stress $\sigma$, there exists a distribution of localised defects that are at (stress) distance $x = \sigma_c - \sigma$ from a stress threshold value $\sigma_c$ at which they will fail. The relevant quantity is the shape of the distribution of the small $x$ values, and we assume an algebraic distribution $P(x \ll 1 ) \sim x^\theta$, with some exponent $\theta$. If the system size is denoted by $N$ and
the density of defects is finite, an extreme value statistics argument determines the minimal value of $x$: it is given by $x_{\rm min} \sim N^{-\a}$, with $\a = 1/(1+\theta)$. Thus, each time the stress is increased by an amount $x_{\rm min}$, a new plastic defect is excited and an avalanche occurs. Now, supposing that the stress is increased by a fixed amount $\Delta \sigma$, this triggers a number $M \sim \Delta\sigma/x_{\rm min}\sim \Delta\sigma N^{\a}$ of independent avalanches. If each avalanche extends over a region of volume $S$, it induces a strain $\delta \gamma \sim S/N$. The total relaxed strain is thus 
\beq
\Delta\gamma \sim M \langle \delta\gamma \rangle \sim M \langle S\rangle/N\sim \Delta \sigma  \langle S\rangle N^{\a-1} \ .
\eeq
We conclude that the average avalanche size is 
\beq\label{eq:2}
\langle S \rangle \sim N^{1-\a} \frac{\Delta\gamma}{\Delta\sigma} \sim N^{1-\a}/\mu_{\rm eff} \ , 
\eeq
where 
$\mu_{\rm eff} = \frac{\Delta\sigma}{\Delta\gamma}$ is the effective elastic modulus that describes the macroscopic stress-strain curve. In addition, it is robustly observed in simulations that avalanches have a scale-free distribution with exponent $\tau$ and a cutoff $S_c$, leading to the functional form
\beq
P(S) \sim S^{-\tau} f(S/S_c) \ , \qquad S_c \sim N^{d_f/d} \ ,
\eeq
where $d_f$ is the fractal dimension of avalanches. Because it is empirically found that $1\leq\tau<2$, we have $\langle S \rangle \sim S_c^{2-\tau} \sim N^{(2-\tau)d_f/d}$, and comparing this with Eq.~\eqref{eq:2} leads to the scaling relation between exponents
\beq
(2-\tau)\frac{d_f}d = \a -1 = \frac{\theta}{1+\theta} \ .
\eeq
The existence of scale-free avalanches has been confirmed, and these scaling relations have been tested both in coarse-grained elasto-plastic models~\cite{lin2014scaling,lin2014density,lin2016mean} and in atomistic molecular dynamics simulations~\cite{franz2017mean,shang2020elastic,franz2021surfing}. Recall that this discussion concerns the pre-yielding regime: scale-free avalanches are also observed in the post-yielding regime. 

Molecular dynamics simulations indicate a universal exponent $\tau=1$, which appears in agreement with mean-field theory~\cite{franz2017mean,shang2020elastic}. The values of $\theta$ and $d_f$ are instead dependent on the glass preparation, with $\theta$ ranging from $\theta \approx 0.1$ for ultrastable glasses to $\theta \approx 0.5$ for the least stable glasses~\cite{ozawa2018random,shang2020elastic}. It has however been proposed that this dependence on glass stability could originate from finite size effects~\cite{lerner2018protocol}, if the distribution $P(x)$ is of the form  $P(x) \sim c x^\theta$ with a strongly stability-dependent prefactor $c$, as suggested by the results of~\cite{wang2019low,rainone2020pinching}. The exponent $\theta$ also has a non-trivial dependence on the applied strain $\gamma$~\cite{hentschel2015stochastic,lin2016mean,ozawa2018random}.

From a theoretical perspective, the agreement between numerical simulations and mean-field theory for the exponent $\tau=1$ is not obvious. In fact, the mean-field prediction is based on the existence of a Gardner phase~\cite{charbonneau2014fractal,franz2017mean} with associated extended excitations leading to extended avalanches, while in numerics avalanches seem to be triggered by highly localized defects that interact elastically~\cite{shang2020elastic}. Whether a more refined mean-field theory of elastically interacting defects can be formulated remains an open problem~\cite{kuhn1997random,das2020robustness,bouchbinder2020low,rainone2021mean}. A first-principle theoretical derivation of the exponent $\theta$, which would clarify whether this exponent is universal or protocol-dependent, is also an important challenge for future research~\cite{muller2015marginal,lin2016mean,ji2019theory}.

\subsection{Memory and training via oscillatory strain}

The complex nature of plasticity in the pre-yielding regime is associated with an underlying rough energy landscape, which thus leads to hysteresis and memory effects. Figure \ref{fig:oscillatory} illustrates this phenomenology for a very stable glass prepared via the swap Monte Carlo algorithm~\cite{swap} in absence of strain. In a single strain cycle, hysteresis is observed provided the amplitude $\gamma_{\rm max}$ is neither too small nor too large. Several numerical simulations have studied a slightly different setting, in which the material is initialized in a random state (for instance by rapid quenching from infinite temperature) and is then trained by repeated application of the same oscillatory strain~\cite{keim2011generic,keim2013multiple,paulsen2014multiple,fiocco2014encoding,fiocco2015memory,mungan2019networks,pashine2019directed,hexner2020effect,keim2019memory}.

As discussed in Sec.~\ref{sec:reversibility}, there is a critical value $\gamma_Y$ beyond which the system displays chaotic, irreversible behavior~\cite{pine2005chaos,corte2008random,fiocco2013oscillatory,regev2013onset,regev2015reversibility,kawasaki2016macroscopic,jin2018stability,bhaumik2019role,yeh2020glass}. The numerical simulations of Refs.~\cite{keim2011generic,keim2013multiple} and the experiment of Ref.~\cite{paulsen2014multiple} trained an emulsion by starting from a random initial state and performing repeated strain cycles of amplitude $\gamma_{\rm train} < \gamma_Y$. After a certain number of cycles, the system settles into a reversible state, akin to that shown in Fig.~\ref{fig:oscillatory}A, in which the configurations visited after each cycle are identical. After training, a read-out experiment is performed, in which the system is subject to a single cycle of variable maximal amplitude $\gamma_{\rm read}$ (Fig.~\ref{fig:oscillatory}D). It is found that for an incomplete training, the system is never fully reversible, but the fraction of moving particles between two subsequent cycles displays a kink when $\gamma_{\rm read} = \gamma_{\rm train}$, i.e. the system shows a memory of the training. For a very large number of training cycle, i.e. when training is complete, no particle moves for $\gamma_{\rm read} < \gamma_{\rm train}$, while motion is observed for $\gamma_{\rm read} > \gamma_{\rm train}$.

The authors then tried alternating training cycles with two (or multiple) values of $\gamma_{\rm train}$. They found that at intermediate training cycles, memories are associated to each of the training strain, but after many training cycles only the largest one persists, because no motion is observed for all $\gamma_{\rm read} < \gamma_{\rm train}$ (Fig.~\ref{fig:oscillatory}E). However, introducing some noise in between training cycles allows the memories to persist indefinitely, see~\cite{keim2011generic,keim2013multiple,paulsen2014multiple} for details.

Similar numerical simulations were performed in \cite{fiocco2013oscillatory,fiocco2014encoding,fiocco2015memory} but in a model for a structural glass, which thus features long-range interactions (Fig.~\ref{fig:oscillatory}F,G,H). It was found that, even after a very large number of training cycles, the system keeps a perfect memory of the training strain, in the sense that no motion is observed during read-out only if $\gamma_{\rm read} = \gamma_{\rm train}$, while some motion is observed if $\gamma_{\rm read} \neq \gamma_{\rm train}$. Thanks to this, multiple memories could be stored even in absence of noise.

From the theoretical point of view, these observation have still to be explained, either within theories of the potential energy landscape, or via simpler effective models~\cite{lindeman2021multiple,hagh2022competition}, which is currently an active research area. 

\section{The yielding transition}

\label{sec:yielding}

\begin{figure*}[t]
\includegraphics[width=0.8\textwidth]{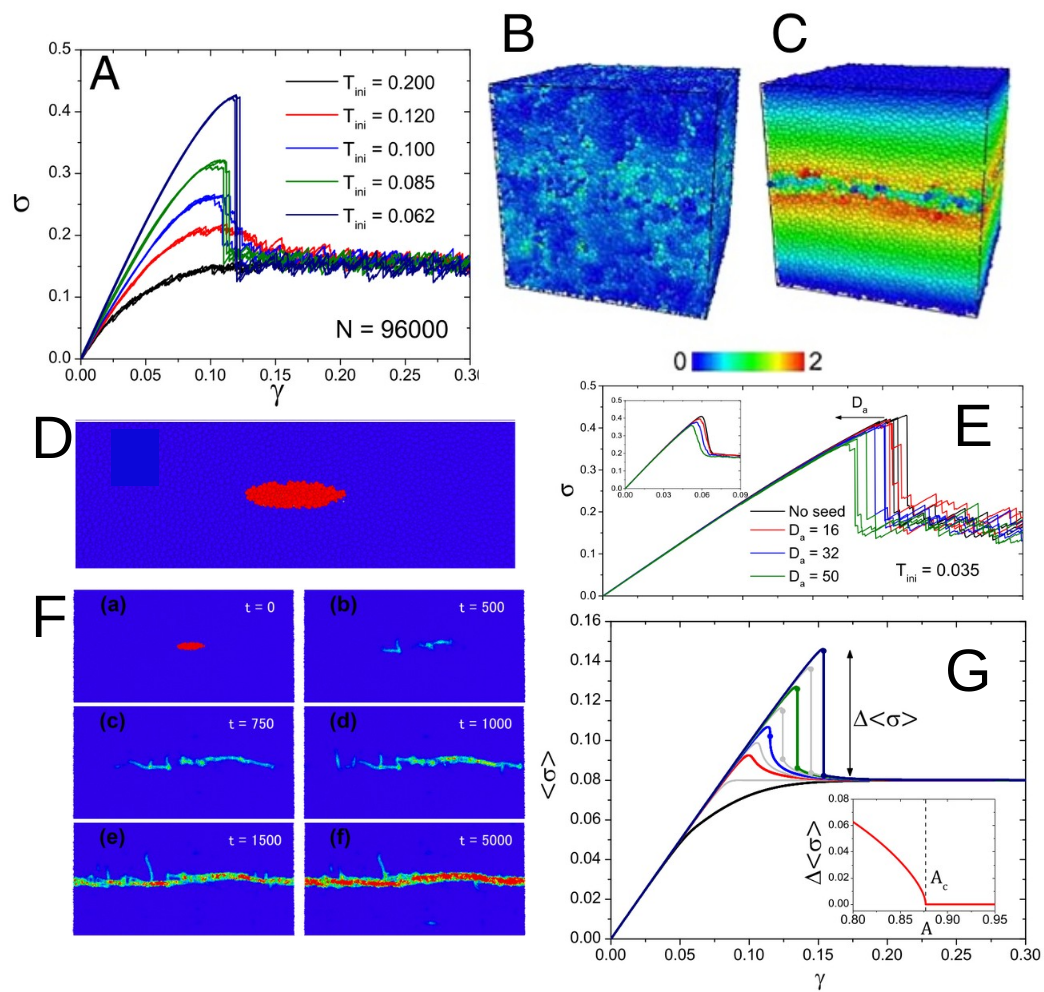}
\caption{{\bf The yielding transition.}
(A) From ductile to brittle yielding in atomistic simulations, from Ref.~\cite{ozawa2018random}. Stress versus strain curves are obtained for different preparation temperatures $T_{\rm ini}$ ($N$ denotes the number of particles used). For reference, $T_g \approx 0.072$ for this system. 
(B,C) Snapshots showing non-affine displacements between $\gamma=0$ and yielding at $\gamma=0.13$ for $T_{\rm ini}=0.120$ and at $\gamma=0.119$ for $T_{\rm ini}=0.062$. (D) Influence of rare defects on brittle yielding, from Ref~\cite{ozawa2021rare}. The figure shows the example of an elongated soft seeding region. (E) Effect of an elongated soft seed of size $D_a \times D_b$ on the stress vs. strain curves of $2d$ glass samples with $N=64000$ atoms and fixed $D_b=8$. Three independent realisations for each $D_a$ are shown for a stable glass with $T_{\rm ini}=0.035$. (F) How shear-bands form, from Ref~\cite{ozawa2021rare}. Time evolution of non-affine squared displacement between $t=0$ and various times $t$ during the gradient-descent dynamics exactly at the yielding transition. At $t=0$, particles inside the seed are colored in red.
(G) Ductile and brittle yielding in the mean-field approximation of the elastoplastic model, from Ref.~\cite{ozawa2018random}. Stress $\langle \sigma \rangle$ versus strain $\gamma$ for increasing degree of annealing ($A$ plays the role of the preparation temperature). The monotonic flow curve (black) transforms into a smooth stress overshoot (red), and above a critical point with infinite slope (blue) becomes a discontinuous transition (green) of increasing amplitude (dark blue). Inset: Stress discontinuity $\Delta\langle \sigma \rangle$ versus the degree of annealing changes continuously across a non-equilibrium phase transition.}
\label{fig:yielding1}
\label{fig:yielding2}
\label{fig:yielding3}
\label{fig:yielding4}
\end{figure*}

Slowly deformed amorphous solids do not all yield in the same way, as shown in Fig.~\ref{fig:intro}. Whereas ductile materials reach a stationary plastic flow through a continuous evolution under applied deformation, brittle ones undergo a macroscopic failure at which the stress discontinuously drops via the formation of a system spanning shear band. 

This phenomenon is obviously of great practical interest, as it broadly relates to how solid materials fail under mechanical deformation. Developing a theory for yielding is also a fundamental challenge for non-equilibrium statistical physics. Recently, a synergy between advanced atomistic simulations, thorough analysis of mesoscopic lattice models, and new theoretical frameworks led to substantial progress on this topic. This section will review the intense research activity that has been taking place in the last few years.

\subsection{Atomistic simulations: stability controls the nature of the yielding transition}

Thanks to the swap Monte Carlo algorithm~\cite{swap}, it has become possible to prepare amorphous solids with very different degrees of stability. This opened the way to a thorough study of the effect of the annealing rate on the yielding transition. The authors of Ref.~\cite{ozawa2018random} used a size-polydisperse model with a soft repulsive potential~\cite{swap}, and by leveraging the swap Monte Carlo algorithm, prepared glass samples with a wide range of stabilities. They used a procedure in which the preparation temperature $T_{\rm ini}$ uniquely controls the glass stability. The range of  $T_{\rm ini}$ was taken to encompass very poorly annealed glasses ($T_{\rm ini} \approx 0.2$, corresponding to wet foam experiments), ordinary computer glasses ($T_{\rm ini} \approx 0.12$, corresponding to colloidal experiments), well-annealed glasses ($T_{\rm ini} \approx 0.085 - 0.075$, corresponding to metallic-glass experiments), as well as  ultrastable glasses ($T_{\rm ini} \approx 0.062$). These preparation temperatures should be compared to the estimated experimental glass transition temperature for this system, $T_g \approx 0.072$. 

The corresponding yielding behavior obtained from strain-controlled athermal quasi-static shear (AQS) deformation using Lees-Edwards boundary conditions~\cite{maloney2004subextensive} is shown in Fig.~\ref{fig:yielding1}A.
For poorly annealed samples, the stress vs strain curve is a monotonously increasing curve. Although there are tiny discontinuous stress drops along the trajectory, the size of those small stress drops decreases with system size, leading to a smooth curve in the thermodynamic limit. The behavior for very stable samples is instead quite different. Yielding is abrupt and associated with a large stress drop, which takes place at a well-defined value of~$\gamma$. This behavior becomes sharper and better resolved as the system size increases, thus signalling a bona-fide discontinuous transition in the thermodynamic limit. Samples prepared with intermediate annealing rates either show a smaller, but still discontinuous, stress drop or a smooth overshoot.

These results have been confirmed by simulations of elastoplastic models~\cite{popovic2018elastoplastic,barlow2020ductile,fielding2021yielding,rossi2022finite} and subsequent atomistic simulations both in two and three dimensions~\cite{ozawa2020role,richard2021brittle}. They show, remarkably, that within a single model of amorphous solid, it is possible to capture the entire range of yielding behaviors found in experiments. The main conclusion of this body of works is that the stability of the amorphous material is the key control parameter of the nature of the yielding transition. Out of the very many microscopic differences distinguishing foams, colloids and molecular glasses, the main parameter controlling the yielding behavior is the different microscopic timescale, which ultimately leads to a different stability of the associated amorphous solid states.  

Furthermore, these findings have important theoretical implications on the statistical physics analysis of yielding in terms of phase transitions. In fact, given that two qualitatively different stress vs strain curves are found by simply changing the value of one control parameter ($T_{\rm ini}/T_g$), 
one expects the existence of a phase transition separating brittle from ductile yielding. The current theoretical understanding of this new out of equilibrium critical point is reviewed in the following subsection. 

\subsection{Theoretical approaches: A critical point separating brittle and ductile yielding}

From the theoretical viewpoint, the first detailed results concerning the nature of the yielding transition were obtained using the mean-field theory of glasses \cite{parisi2020theory}. By formally following the evolution of the free-energy landscape of the system as it is gradually deformed, the yielding transition can be described as a discontinuous transition displaying the same critical property as a spinodal instability~\cite{wisitsorasak2012strength,rainone2015following,parisi2017shear}. The structural heterogeneity of amorphous solids then introduces quenched disorder, as it leads to spatial fluctuations in the local degree of stability, leading to regions that are more prone to rearrange plastically than others. Therefore, this approach suggests that yielding should be treated as a spinodal instability in the presence of quenched disorder. This problem has been studied in more detail in the context of the Random Field Ising Model (RFIM)~\cite{nattermann1998theory}, which describes a ferromagnetic material subjected to local quenched disorder. In that case, the spinodal instability is observed when the external magnetic field is quasi-statically varied starting from a magnetized configuration at zero temperature. For the RFIM, finite-dimensional fluctuations bring important new ingredients compared to the mean-field description, and can change the nature of the spinodal transition, which is no longer critical but is instead governed by rare fluctuations~\cite{nandi2016spinodals}. Similarly to yielding~\cite{ozawa2018random}, the RFIM also displays for strong disorder a smooth magnetization vs magnetic field curve, which becomes instead discontinuous at smaller disorder. A critical point, associated to a non-equilibrium phase transition, separates these two regimes~\cite{perkovic1995avalanches}.

There are many similarities between the RFIM spinodal and the yielding transition of amorphous solids, but also important differences. The most important one is that the interaction between a plastically rearranging region and the rest of the system is not mediated by a short-range and positive kernel but by a long-range and anisotropic one, due to the elastic deformation of the material. As a consequence, whether the phase transitions found for the RFIM hold mutatis mutandis for the yielding of amorphous solids needs scrutiny. 

One of the main issues investigated recently is the existence of a bona-fine critical point 
separating brittle and ductile regimes. The atomistic simulations of \cite{ozawa2018random} found direct evidence of such phase transition by identifying diverging susceptibilities accompanying the transition from brittle to ductile yielding. However, subsequent theoretical work~\cite{barlow2020ductile,fielding2021yielding} and large-scale atomistic simulations~\cite{richard2021finite} questioned the resulting phase diagram, suggesting that no ductile phase exists and that yielding is always discontinuous for large enough systems. It is however difficult to reach a firm conclusion due to the limited system sizes and small number of samples accessible in the atomistic simulations.

This issue can instead be settled using elasto-plastic models. The non-equilibrium critical point was studied by mean-field approximations in Refs.~\cite{ozawa2018random,popovic2018elastoplastic} and by numerical simulations in Refs.~\cite{popovic2018elastoplastic,rossi2022finite}. In these models, the degree of annealing can be represented through the initial distribution of the local stress (narrow for well-annealed systems, broad for poorly annealed ones)~\cite{baret2002extremal}. 
The mean-field analysis supports the existence of a phase transition separating brittle and ductile yielding. As an example, we show in Fig.~\ref{fig:yielding2}G the stress vs strain curve obtained in \cite{ozawa2018random}. It is however important to study the effect of finite-dimensional fluctuations, in view of the results of Refs.~\cite{barlow2020ductile,fielding2021yielding}. The large-scale numerical simulations of two and three dimensional elasto-plastic models found the same finite-size effects highlighted in~\cite{richard2021finite}, but were also able to show that the critical point separating brittle and ductile behaviors persists in the thermodynamic limit. Thanks to the coarse-grained lattice nature of elasto-plastic model, one can reach sizes that are roughly 100 times larger than the ones in atomistic simulations. The existence of a smooth overshoot in the stress vs strain curves in the thermodynamic limit is however not yet fully settled.  

\subsection{Brittle yielding: shear-bands and the role of rare fluctuations}

The discontinuous spinodal transition of the RFIM is governed by rare regions that act as seeds for the macroscopic avalanche associated to the discontinuous jump of the magnetization \cite{nandi2016spinodals}. The strong analogy between yielding and the physics of the RFIM suggests that a similar mechanism may be at play for brittle yielding, as proposed and investigated in Refs.~\cite{ozawa2018random,popovic2018elastoplastic,ozawa2021rare}.

As shown in Fig.~\ref{fig:yielding1}C, brittle yielding is associated to the formation of a macroscopic shear-band, which can be interpreted as a macroscopic avalanche. It is natural to expect that within a stable solid there exists a very small concentration of weak spots that are more prone to rearrange plastically than the rest of the material. Such soft regions are created by density fluctuations that are frozen-in during the formation of the amorphous solid. These regions can rearrange (possibly multiple times) before the bulk becomes unstable. As a consequence, they can act as nucleation seeds for the formation and propagation of a shear-band. This picture is supported by theoretical arguments based on the formation of aligned Eshelby quadrupoles within the seed region, and considerations based on a generalization of fracture theory~\cite{popovic2018elastoplastic,ozawa2021rare}. 

Atomistic simulations are unable to directly probe this phenomenon, as only very small soft regions are found even in the largest systems that can be studied numerically. To circumvent this problem, the authors of \cite{ozawa2021rare} inserted a soft region in an otherwise stable glass (see also \cite{ozawa2018random,popovic2018elastoplastic} for previous results and related investigations in elastoplastic models). They did this by preparing the stable glass first, and then annealing by Monte Carlo simulations a small region of space that will form the soft seed. Confirming the above arguments, it was found that the presence of a soft seed considerably facilitates the failure of the material, decreasing the value of the yield strain, an effect which becomes more important for larger seeds, see Fig.~\ref{fig:yielding3}E. These simulations confirm that under applied deformation the soft seed region relaxes plastically much before the bulk. This relaxation then destabilizes the surrounding particles, which then also yield before the bulk. This leads to a growth of the soft region and, beyond a certain value of the strain, to a self-sustained process eventually leading to the formation of a macroscopic shear-band. These simulations show that rare soft regions indeed act as nucleation seeds for shear-bands (Fig.~\ref{fig:yielding4}F) which eventually produce a macroscopic stress drop.

In summary, brittle yielding is controlled by rare events that cause the nucleation and propagation of shear-bands. This phenomenon shares similarities with fracture~\cite{popovic2018elastoplastic} but also important differences, as the soft regions are not empty voids but represent instead spontaneous structural fluctuations frozen at the glass transition in macroscopic samples. 

\begin{figure*}[t]
\includegraphics[width=\textwidth]{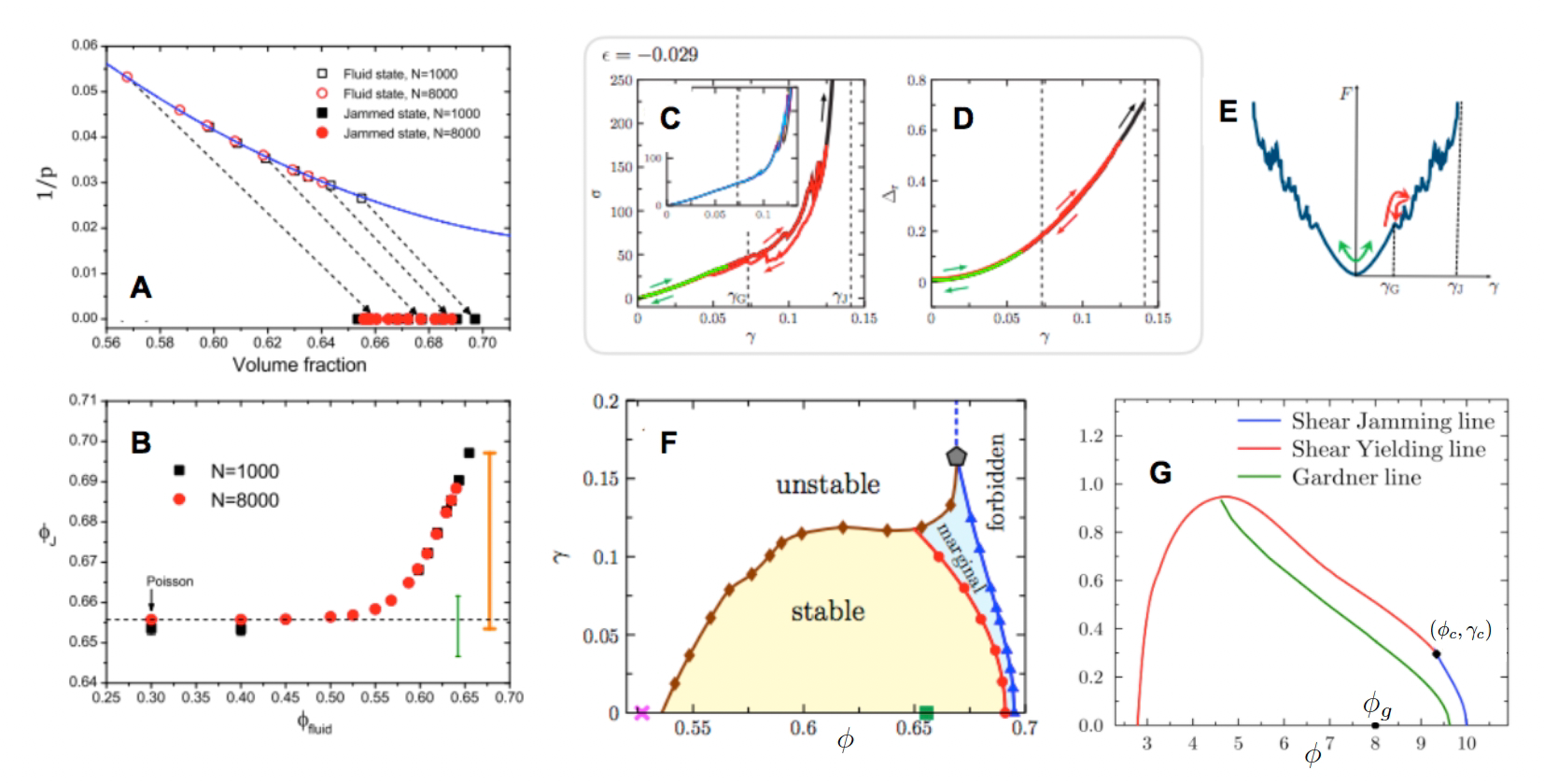}
\caption{
{\bf Shear yielding and shear jamming.}
(A) Compression of a hard colloidal glass in equilibrium up to an initial density $\phi_{\rm ini}$, followed by rapid compression up to jamming, where the pressure $p$ diverges, and
(B) the final jamming packing fraction $\phi_J$ as a function of the initial $\phi_{\rm ini}=\phi_{\rm fluid}$, both from Ref.~\cite{ozawa2017exploring}.
(C) A strain $\gamma$ is applied adiabatically on a stable glass prepared at
$\phi_{\rm ini}=0.655$, and compressed at $\phi=0.675$ (as in Fig.~\ref{fig:oscillatory}A, which is for the same glass at lower $\phi=0.66$), from Ref.~\cite{jin2018stability}. Here the stress diverges at a finite strain, indicating shear jamming.
(D) Displacement with respect to the unstrained configuration, demonstrating reversibility, and
(E) sketch of the corresponding free energy landscape, both from Ref.~\cite{jin2018stability}.
(F) Phase diagram for a glass prepared at $\phi_{\rm ini}=0.655$, then compressed, decompressed or strained in the $(\phi, \gamma)$ plane. For low $\phi$, shear yielding is observed, while at high $\phi$, shear jamming happens, from Ref.~\cite{jin2018stability}.
(G) Same phase diagram computed within mean-field theory at $d\to\infty$, from Ref.~\cite{altieri2019mean}.\\
}
\label{fig:jamming}
\end{figure*}
\subsection{Perspectives and open questions on the yielding transition}

The intense research effort of the last years on the yielding transition led to substantial progress, but several important questions remain open. We suggest four lines of investigations for the future. 

First, the way in which the material rearranges plastically in real space is very different for ductile and brittle yielding, as shown in Figs.~\ref{fig:yielding1}B,C. An interesting point both for applications and theory is to understand how the two are connected, i.e. how the nature and the role of shear-bands change when going from brittle to ductile yielding.  

Second, although the existence of a ductile and a brittle phase separated by a critical point is to a large extent supported by simulations and theory, whether a smooth stress overshoot can exist in the stress versus strain curves in the thermodynamic limit is an open and debated question~\cite{barlow2020ductile,rossi2022finite}. 

Third, although the analogy between yielding and the RFIM spinodal has been a useful guideline, establishing to what extent there is a strong connection remains an open problem. On the one hand, it is clear that the short-range ferromagnetic Ising model cannot be used to describe quantitatively the critical properties associated to yielding, as the anisotropic and long-range nature of the elastic interaction matter for physical properties such as shear-bands and avalanches. This has been fully investigated in mean-field models \cite{parley2023towards,rossi2022emergence}. Moreover, the fact that spins can only flip once, whereas mesoscopic regions can fail multiple times is an important difference between the two systems. On the other hand, the Random Field Ising Model with Eshelby-like interactions could be an effective model displaying the same critical properties of  the transition separating brittle and ductile behaviors \cite{rossi2023far}.

Finally, the recent results described above were obtained in the idealized quasi-static and zero-temperature limits. Understanding quantitatively how the physics changes in the presence of small but finite shear rates~\cite{singh2020brittle} and temperatures is an important issue to be addressed in order to compare theoretical results with experiments. Several works~\cite{popovic2021thermally,popovic2022scaling,sollich2017aging,parley2020aging} have recently attempted to include thermal fluctuations in the context of elasto-plastic models. These could serve as useful starting points to study the effects of thermal fluctuations on the yielding transition itself.  

\section{Rheology near jamming}

So far, we have considered thermal systems with soft and smooth repulsive interactions. We considered glassy states prepared by equilibrium preparation at some initial temperature $T_{\rm ini}$, followed by a rapid quench to zero temperature, and finally subjected to an applied deformation.

However, several interesting phenomena in the rheology of amorphous solids are related to the presence of a strong hard core repulsion (or equivalently to a finite range repulsive potential at zero temperature), which leads to the existence of a jamming transition \cite{liu2010jamming}. The analog of thermal cooling is, for a colloidal hard sphere glass, a slow compression that maintains the hard sphere system in equilibrium up to an initial packing fraction $\phi_{\rm ini}$, followed by a rapid compression to the jamming point $\phi_J$ where pressure diverges and particles remain mechanically blocked by the hard cores~\cite{ozawa2012jamming,ozawa2017exploring,charbonneau2021memory} (see Fig.~\ref{fig:jamming}A for the thermodynamic path). By analogy with the potential energy of the inherent state in the thermal case, the jamming density is found to increase upon increasing the initial density (Fig.~\ref{fig:jamming}B), indicating increased stability of the resulting hard sphere glass.

The jamming transition controls the formation of rigid glassy states upon compression, e.g. in emulsions, colloids, or granular materials. Such materials display the phenomenon of dilatancy, wherein their volume increases upon constant-pressure shear deformation, or similarly their pressure increases upon constant-volume deformation.

Based on observations in steady flow, it has long been thought that dilatancy is intrinsically associated with friction~\cite{peyneau2008frictionless,bi2011jamming,seto2019shear}. However, recent results - both from analytical mean-field theory~\cite{rainone2015following,urbani2017shear,altieri2019mean} and from numerical simulations~\cite{vinutha2016disentangling,jin2018stability} - demonstrated that this phenomenon also occurs in frictionless sphere packings in the transient start-up shear regime, provided the system is prepared in a stable enough glass state.

In an extreme version of dilatancy, the pressure can increase so much that it diverges upon shearing at constant volume, which leads to shear-jamming at a finite value of the strain (Fig.~\ref{fig:jamming}C shows the stress as a function of strain, with the pressure being roughly proportional to the shear stress)~\cite{rainone2015following,urbani2017shear,vinutha2016disentangling,jin2018stability}. The material is then brought into a jammed state by the application of a strain, and it supports an infinite stress which prevents mechanical failure (yielding). This happens while the system remains confined within a specific glass basin in the free-energy landscape (Fig.~\ref{fig:jamming}D,E), leading also to the partially reversible regime defined above. These results demonstrate again that, depending on glass stability, either shear-jamming or shear-yielding can be observed in hard-core particle amorphous assemblies, leading to a non-trivial phase diagram for a glass prepared at fixed initial density $\phi_{\rm ini}$ and then compressed/decompressed and strained to a state point $(\phi,\gamma)$ (see Fig.~\ref{fig:jamming}F). The existence of the solid is bounded by the shear-yielding line at low $\phi$ and by the shear-jamming line at high $\phi$, which are separated by a critical point. Qualitatively similar results have been obtained within mean-field theory in $d\to\infty$ (Fig.~\ref{fig:jamming}G)~\cite{urbani2017shear,altieri2019mean}.

These results show in particular that while glasses of different initial density display equivalent properties under isotropic compression, they exhibit striking differences in rheological behaviour under shear. These results have been further applied to analyse the behaviour of non-Brownian soft spheres, resulting in a particularly rich and complex phase diagram~\cite{jin2021jamming,babu2021dilatancy,PAN20231}.

\section{Plasticity defects}
\label{sec:defects}

In order to quantitatively connect continuum theories for pre-yielding behavior and the yielding transition to specific features of atomistic simulations and experiments, we need to identify the microscopic regions that can yield within a glass. In crystalline solids, a microscopic theory of deformation begins with topological defects in the otherwise perfectly periodic crystalline structure, such as dislocations. Until fairly recently, it remained unclear whether similar defects exist in glasses, and initial searches focused on obvious structural quantities like local free volume~\cite{sastry1998freevolume} did not correlate strongly with yielding and plasticity. 

Instead, some older~\cite{schober1996low} and more recent numerical work~\cite{widmercooper2006predicting, tanguy2010vibrational, manning2011softspots}, has demonstrated that subtle features in the vibrational spectrum can help identifying defect-like sites within glasses and confirmed that microscopic plastic rearrangements occur at these sites. Since the number of defects --  as well as their stiffness, energy barriers, and interactions -- control the spatio-temporal evolution of avalanches that occur in both the pre-yielding and yielding regimes, a first-principles understanding of defect properties in glasses would significantly strengthen the predictive power of continuum theories.
In this section, we will thus describe a number of distinct ways to identify soft defects that contribute to the material plasticity and yielding.

\begin{figure}[t]
\includegraphics[width=\columnwidth]{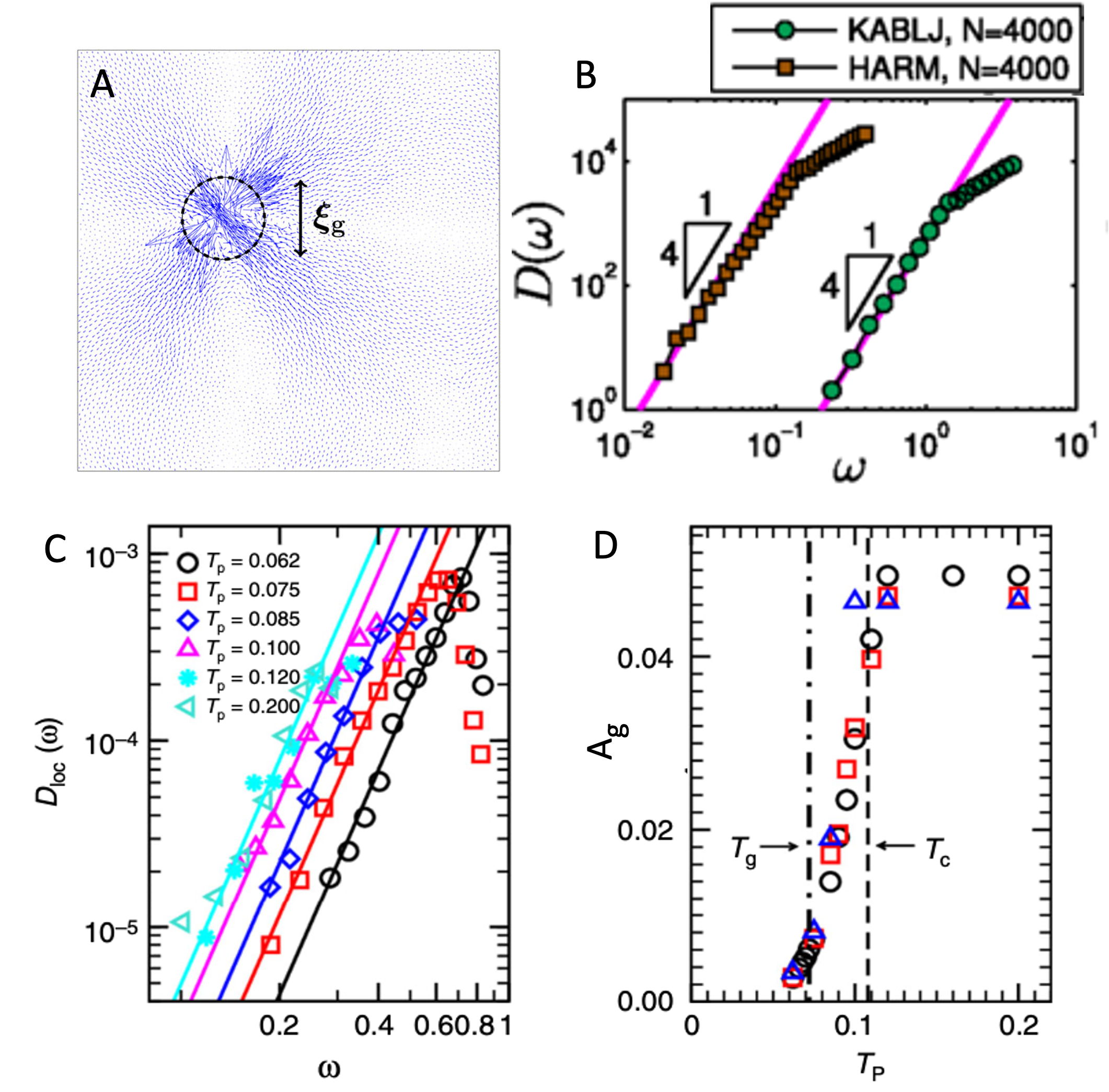}
\caption{
{\bf Plasticity defects in amorphous solids.}
(A) The blue arrows overlaid on a 2D particle packing show an example of a quasi-localized mode in the vibrational spectrum of this solid. 
(B) Density of vibrational modes in a particle packing with Lennard-Jones (KABLJ) and harmonic (HARM) interaction potentials, illustrating the $\omega^s$ dependence with $s=4$. 
(C,D) Density of states for systems prepared using swap Monte Carlo at different initial temperatures $T_{ini}$, ranging from brittle to ductile glasses. While the $\omega^4$ regime exists in all these glasses, the prefactor parameterized by $A_g$ deacreases as $T_{ini}$ decreases, indicating an overall decrease in the density of these modes compared to other modes in the system.}
\label{fig:defects}
\end{figure}

A first way to identify such defects is to look at the harmonic vibrations around a glassy energy minimum (or inherenet structure). It has been shown that such minima feature a universal band of quasi-localized modes in the density of vibrational states, whose density of states scales as $g(\omega) \sim A_g \omega^s$ with an exponent $s=4$ in most cases~\cite{lerner2016statistics,richard2020universality}.
Although the precise values of $s$ is a matter of debate~\cite{wang2023scaling, lerner2021low,schirmacher2023nature}, it does not impact the relevance of these modes for plasticity, which is our main focus here.
An example of the density of states for amorphous solids with two different interaction potentials (harmonic and Kob-Andersen Lennard Jones) is shown in Fig~\ref{fig:defects}B.
These low-frequency modes correspond to quasi-localized excitations (QLE), composed of a disordered, localized core surrounded by a four-fold symmetric long-range elastic field similar to that predicted by Eshelby~\cite{eshelby1957determination}, an example of which is shown in Fig~\ref{fig:defects}A.

The prefactor $A_g$ for the $\omega^s$ scaling regime can be understood as the product of the average stiffness, quantified by a characteristic oscillation frequency $\omega_g$, of the QLEs and the total number of such excitations in the material $\cal{N}$, with ${\cal N} = \int_0^{\omega_g} {\rm d} \omega A_g \omega^s$~\cite{lerner2016statistics,kapteijns2018universal,richard2020universality, lerner2021low}. Fig~\ref{fig:defects}C shows that the density of localized modes in simulations with different preparation temperatures $T_{ini}$ (prepared using the swap Monte Carlo algorithm) have systematically different prefactors $A_g$, which corresponds to the offset of the intercept on these log-log plots. Fig~\ref{fig:defects}D represents the evolution of $A_g$ with $T_{\rm ini}$ highlighting that more stable glasses have a much smaller prefactor $A_g$, and therefore have fewer and stiffer defects~\cite{wang2019low}. The prefactor $A_g$ changes most rapidly near the mode-coupling crossover temperature.

What is the physical origin of such low-frequency quasi-localized modes? A few works have studied random networks or random potentials that can generate such modes, though those require some fine-tuning or are based on strong assumptions. Very recently, a scaling theory for mean-field interacting anharmonic oscillators has been developed that may explain some features of this vibrational mode regime~\cite{das2020robustness,rainone2021mean,bouchbinder2020low}. The theory focuses on three parameters: a cutoff scale associated with the harmonic stiffness $\kappa_0$ of the oscillators, the typical strength of the random couplings between oscillators, $J$, and the strength of their interaction with a  surrounding elastic medium, $h$. In this model, the $\omega^4$ scaling exists across a wide range of parameters, and there exists a weak coupling regime where the prefactor $A_g$ varies exponentially with the quantity $-\kappa_0 h^{2/3}/J^2$, which is reminiscent of the exponential variation of $A_g$ with the inverse of the parent temperature that characterizes the preparation protocol seen in numerical simulations.
The alternative theoretical framework of Heterogeneous Elasticity Theory~\cite{schirmacher2023nature} provides similar predictions, but with an exponent $s$ that depends on the microscopic details.
Open questions include understanding how these theories can be directly connected to more realistic models for interactions between defects, and identifying new methods that can separately analyse the density and the stiffness of defects as for now only the product is accessible from the density of states.

A second interesting class of defects that are known to exist in glasses are tunnelling two-level systems (TLS), which are localised excitations that give rise to a universal and anomalous linear specific heat~\cite{anderson1972anomalous,phillips1987two,reinisch2005moving,damart2018atomistic}.
It has been speculated that TLS and QLE are related~\cite{khomenko2020depletion,ji2019theory},
but while QLEs are harmonic modes with small energy barriers, two-level systems are states where nearby minima in the landscape are very close in energy. It remains unclear if two-level systems can be understood as a special subset of QLEs~\cite{kumar2021density}.

Recent work has vastly improved our ability to identify and characterize defects beyond the harmonic regime of QLE and that of TLS, thus identifying several other classes of defects.
The work most directly related to elasto-plastic models in the previous section is a method to approximate the local yield stress $x$, and its distribution $P(x)$, developed by Patinet and collaborators for 2D glasses~\cite{barbot2018local}, with recent extension to three dimensions~\cite{ruan2022predicting}. In this method, a shear strain is locally applied within a spherical patch of particles, and one measures the amount of shear stress required for the patch to yield. Other methods have identified purely structural signatures of defects using high energy motifs that can be identified in systems with specific interaction potentials~\cite{tong2018revealing} or machine learning approaches~\cite{schoenholz2016structural, bapst2020unveiling}.  Another set of approaches studies nonlinear modes~\cite{gartner2016nonlinear} associated with terms beyond second order in the expansion of the energy in terms of particle displacements, or their approximations~\cite{richard2021simple}. These methods are particularly useful in situations where the defects are stiff (associated with high curvatures in the potential energy landscape) and weaken significantly under shear.

Many of these methods were recently studied together on the same data set over a broad range of material stabilities in the pre-yielding and yielding regimes~\cite{richard2020predicting}. This analysis identified which structural defect indicators were most effective in various situations -- for example, linear response is surprisingly effective in ductile materials, while nonlinear modes and local yield strain are superior in very stable systems. Moreover, all the effective indicators concurred that the initial number of low-energy barrier defects was significantly smaller in stable materials, leading to spatial self-organization into shear-bands at the yielding transition~\cite{richard2020predicting}.

An important question concerns the length scale, $\xi$, characterizing the core of the defects, as highlighted in Fig.~\ref{fig:defects}A. Recent work has indicated that close to the jamming transition, the size of the defect core $\xi$ grows as the pressure decreases, $\xi \sim p^{-1/4}$~\cite{shimada2018spatial}. This scaling relation supports the idea that QLEs are anomalous modes that are related to the boson peak in jammed solids~\cite{lerner2014breakdown, lerner2022boson}, and very near jamming they can become extended. Importantly, avalanches found under shear are also modified around the jamming transition, suggesting that the spatial extent of excitations may significantly alter their interactions~\cite{franz2017mean}. A detailed understanding of the evolution of these localized defects close to the 
jamming transition is lacking~\cite{giannini2023scaling}.

Most of this initial work on defects has focused on systems with relatively simple, spherically symmetric interaction potentials, in the limit of zero temperature and zero strain rate.  It will be very important in the future to quantify how the features of the defect population, as well as interactions between defects, change at finite temperature and strain rates, and with more realistic interaction potentials. 

\section{Plasticity, defects and yielding in biological tissues}

\begin{figure*}[t]
\includegraphics[width=\textwidth]{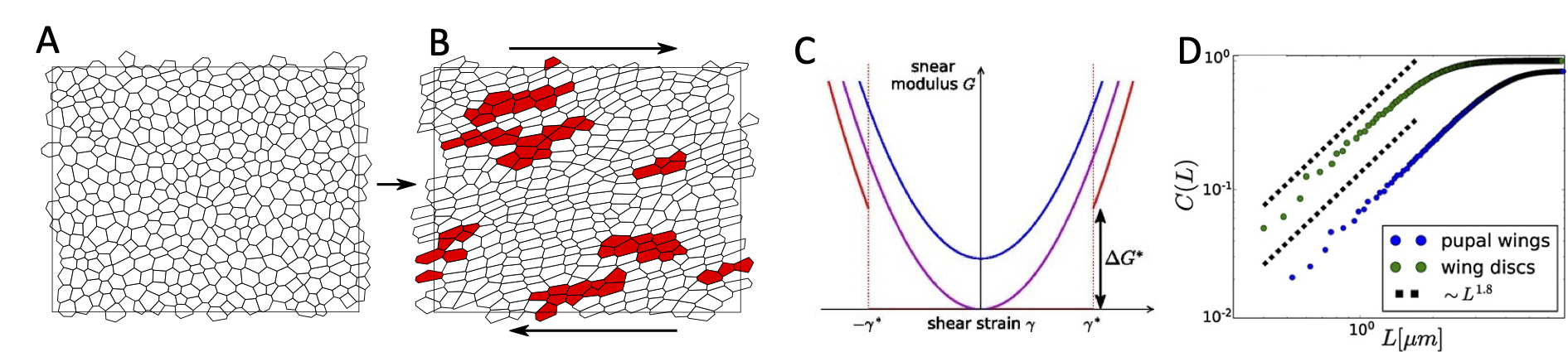}
\caption{
{\bf The vertex model for biological tissues.}
(A,B) Two snapshots representing the effect of shear in the vertex model, with plastic rearrangements highlighted in red, from Ref~\cite{popovic2021inferring}.  
(C) Scaling prediction for the variation of the shear modulus of vertex models with the shear strain for different values of the shape parameter (blue, shape is below critical, magenta at critical point, red is above criticality), from Ref~\cite{merkel2019minimal}. 
(D) Cumulative distribution of distance to yield stress $C(L) = \int P(L)$ in two tissues from the fruit fly embryo, from Ref~\cite{popovic2021inferring}.
}
\label{fig:vertex}
\end{figure*}

There is a growing interest in understanding the rheology of active and biological matter. For amorphous biological materials, such as the tissue that comprises a developing organism or the mechanics of cytoskeletal machinery within the cell, evidence is emerging that the rheology of the biological material enables specific biological processes, such as body axis elongation. For designed amorphous materials like active colloids or synthetic-circuit driven bacterial swarms, the rheology of the self-driven material dictates design principles and can be harnessed to generate novel material behavior. 

These non-quiescent amorphous materials are different in important ways from their condensed matter counterparts described in previous sections. For example, the effective interaction potential between components is different for biological tissues or cytoskeletal networks, while the driving forces are different for active matter. Since this field is broad and expanding rapidly, we will not attempt to provide a comprehensive review. Instead, we will focus here on the yielding transition in the limit of slow driving, with a special emphasis on how the ideas explored in the previous sections can be extended and applied to tissues and active matter.

First, we review features of yielding in biological tissues. We use the term to refer to a broad class of cellularized (cells are packed tightly together with many cell-cell contacts) and acellularized (few cells, mostly composed of sparse collagen or other fiber networks called extra-cellular matrix) materials. At first glance, these materials share many similarities with the glasses discussed above. They are composed of disordered, strongly-interacting units, and their microscopic structure is reminiscent of foams or jammed spheres. However, a major difference is that interactions are often topological, in the sense that they depend on the topology of an underlying interaction network, rather than on the metric interactions that depend on the distance and angles between the components that are more common in soft condensed matter. As a consequence, biological systems may have fewer constraints between their degrees of freedom. Interestingly, confluent cellularized tissues -- where there are no gaps or overlaps between cells -- usually have underconstrained topological interactions, whereas less dense tissues -- where there are large gaps between cells -- appear to have metric interactions. Recent models are able to interpolate between these two possibilities~\cite{kim2021embryonic}.  Furthermore, fluctuations in these materials can be driven either by thermal fluctuations or by several different active biological processes, including tension induced by motor proteins, active cell crawling, and more.

From a biological perspective, it is important to understand the yielding transition because there is growing experimental evidence that organisms exploit this rheological transition during development and disease. For example, in vertebrate embryos, the shear-induced yielding induced by growth of the nascent spine is necessary to allow the body axis to elongate~\cite{mongera2018fluid}.  There are even some simple organisms that seem to use brittle failure mechanisms to reproduce~\cite{prakash2021motility}. Similarly, during early stages of development a yielding transition is required to allow gastrulation (i.e. the formation of the gut tube)~\cite{petridou2021rigidity}, and cancer tumor spheroids lower their yield stress as they become more invasive~\cite{grosser2021cell}, which is important for how cells shear off from the main tumor during metastasis~\cite{ilina2020cell}.

From a physics perspective, it is interesting to understand whether the differences in the nature of the interaction network in these materials alter the nature of the yielding transition.  To tackle this question, note that the nature of the underlying zero-temperature rigidity transition is often different in systems with topological interactions compared to jammed spheres and other materials with metric interactions. It is well established that jamming rigidity occurs precisely when the number of network connections (generated by metric interactions between particles) equals the number of degrees of freedom. This is termed Maxwell-Calladine constraint counting, which arises from considering first-order perturbations to the constraints (e.g., the length between neighboring particles interacting via a two-body potential). In contrast, constraint counting in many rigid biological systems suggests that the systems are always underconstrained, i.e., the number of network connections is less than the degrees of freedom. Recent work has demonstrated that such systems become rigid as a continuous parameter is tuned (cell shape in confluent cellularized tissues, strain in fiber networks)~\cite{bi2015density,merkel2019minimal}, because there are energy penalties that only occur at second-order in perturbations to the constraints~\cite{damavandi2021energetic}. 

These investigations of the nature of the rigidity transition have led to useful rheological predictions based on a scaling theory for how the finite-strain shear modulus of the material changes as a function of the internal tuning parameter~\cite{merkel2019minimal}. Since the scaling is a universal feature in such models, it works for both cellularized tissues and fiber networks. Essentially, it predicts that if the internal tuning parameter is above its critical point (e.g. a cell shape is above its critical shape index) then the material posesses a zero shear modulus up to a critical value $\gamma^{*}$ with a discontinuous jump $\Delta G^{*}$ that is also predicted by the theory, and behaves quadratically thereafter, as shown in Fig~\ref{fig:vertex}C. A recent work developed a more detailed single-cell mean-field approach of this phenomena~\cite{huang2022shear}. In contrast, if the system is below the critical point, the scaling is always quadratic, and the minimum modulus is set by the internal tuning parameter. These predictions have been validated in simulations of tissues and fiber networks. While seemingly consistent with existing experiments, new experiments are needed to test additional predictions of the theory.

An obvious question is whether this difference in the nature of the transition impacts the vibrational spectrum, defects, avalanches, and ultimately the yielding behavior.  While much work remains to be done, there have been some initial studies to address these questions. First, even near the rigidity transition, the vibrational spectrum of a 2D Voronoi model for confluent tissues is quite different from that of jammed packings. It does not exhibit a plateau in the density of states associated with a boson peak in glasses, and no low-frequency band of quasi-localised modes has been identified~\cite{sussman2018anomalous}. Furthermore, the inverse participation ratio remains very low at the lowest frequencies, suggesting that the linear modes at low frequencies remain extended. One may be tempted to assume, therefore, that such materials do not possess the same localized defects as glasses. 

However, simulations of sheared tissue models make it clear that rearrangements do tend to occur in localized patches, as shown in Figs.~\ref{fig:vertex}A,B. Moreover, local structure strongly facilitates rearrangements. Recent machine learning approaches have even been used to identify localized structural signatures that are strongly correlated with future cell rearrangements~\cite{tah2021quantifying}. In addition, a very simple localized structural quantity (having a short cell edge length) has also been shown to be an excellent predictor of future plasticity~\cite{popovic2021inferring}. These results underlie the importance of localized excitations in tissue models.  

The authors of Ref.~\cite{popovic2021inferring} use the existence of a cusp in the energy at a cell rearrangement to develop a simple scaling argument that predicts how the force required to yield (denoted $x$ in the elasto-plastic models discussed above) scales with the length of the short cell edge. Specifically, one can apply a strain that shrinks the length of an edge with an equilibrium length $L$ to a new value $L^*$. Expanding the energy around $L$ gives $\Delta E(L^*) \approx \Delta E(L) + 1/2 \Delta E''(L) (L^*- L)^2$.  At the transition point where $L^* = 0$, this predicts that the energy is $\Delta E (L^* = 0) \approx 1/2 \Delta E''(L) L^2$, and the force on the edge required to trigger the rearrangement is therefore $x \approx \Delta E''(L) L$, suggesting the simple scaling relation $x \sim L$. 

This scaling relation was succesfully tested in numerical simulations of a 2D vertex model~\cite{popovic2021inferring}. This then allows the authors to extract the distribution of short edge lengths in numerical studies and experiments, and extract $P(L) \sim L^\theta$ or $C(L) = \int P(L) \sim L^{\theta + 1}$, see Fig~\ref{fig:vertex}D. In numerical simulations $\theta \approx 0.5-0.6$, while in the developing fruit fly wing one finds $\theta \approx 0.7-0.9$. Since in elasto-plastic models the distribution $P(x)$ completely determines whether the yielding behavior is brittle or ductile, there is hope that this preliminary work may help to characterize the mechanical ductility of tissues, organs, and organisms, thereby allowing the prediction of mechanisms for morphogenesis and metastasis. Moreover, the fact that identifying $P(x)$ directly in topologically interacting systems (such as vertex or Voronoi models) is so much easier than in particulate or metric systems suggests that such models may be an excellent place to carefully test elasto-plastic model predictions.

One may wonder why these structural defects do not show up in the linear spectrum.  One intriguing possibility is that since the rigidity transition itself relies on higher-order perturbations to the constraints, perhaps higher order-terms in the expansion of the energy (beyond the dynamical matrix) are required to find those defects. Is there some universality in the spectrum of higher-order terms?

Another open question is how the yielding transition changes in the presence of finite fluctuations or at finite strain rates. Ongoing work to study the nonlinear rheology of confluent tissue models suggest that they are shear-thinning, yield stress solids~\cite{duclut2021nonlinear, sanematsu20213d}, and that under finite applied strain the fluid phase can rigidify due to geometric effects, much like shear jamming in particle systems~\cite{huang2022shear}. Even the linear rheological behaviour exhibits interesting, nontrivial features at finite frequencies~\cite{tong2022linear}. Recently, a constitutive model for biological tissues deformed at finite strain rates has been proposed to capture such features~\cite{fielding2022constitutive}. At finite temperatures, confluent tissues exhibit anomalous sub-Arrhenius relaxation dynamics, with effective energy barriers that appear to become smaller as the temperature decreases~\cite{sussman2018anomalous}. Clearly, more work is needed to fully understand similarities and differences between biologically relevant materials and physical glassy systems. 

\section{Rheology, plasticity and yielding in active matter}

\begin{figure*}[t]
\includegraphics[width=.9\textwidth]{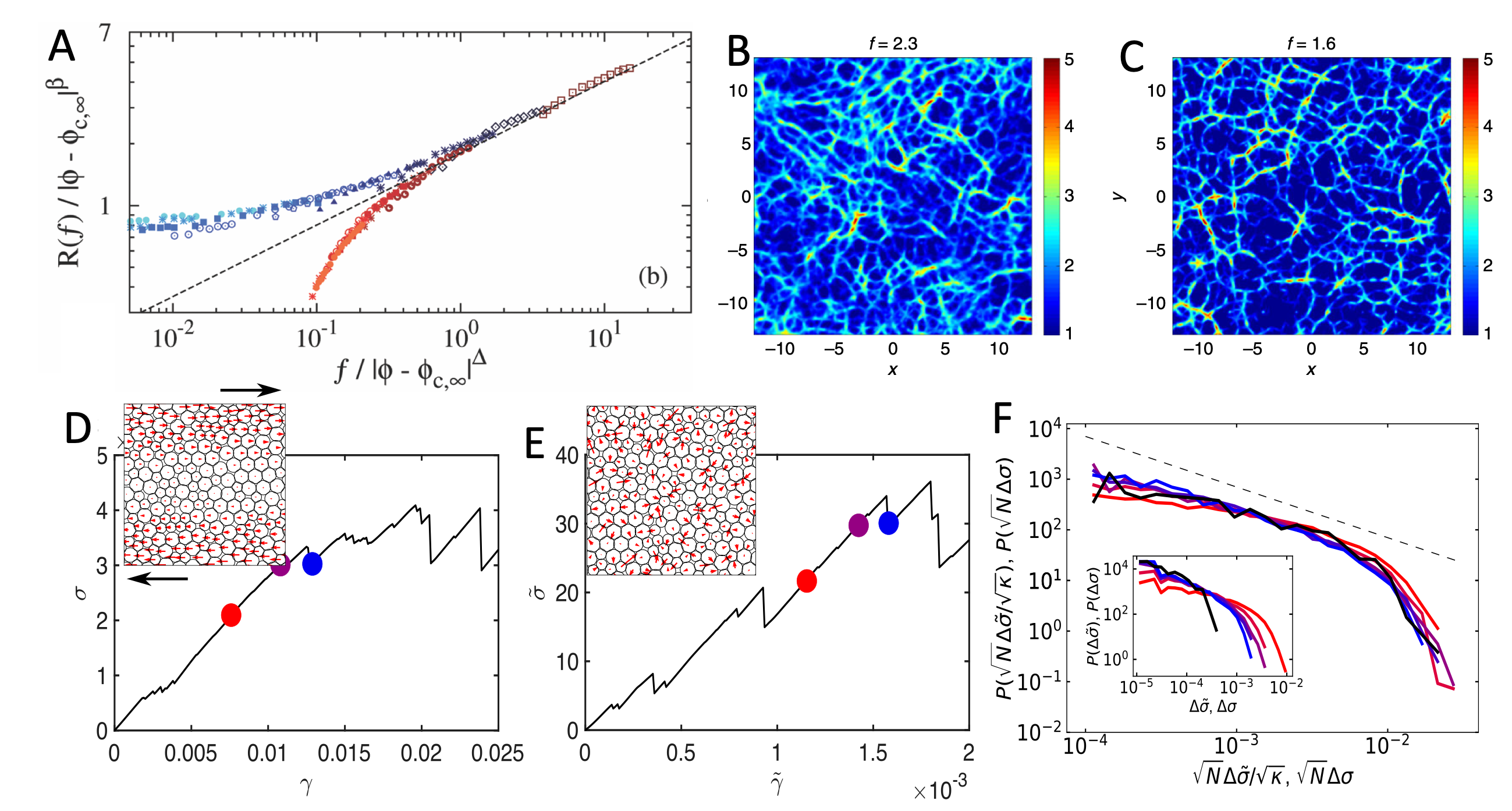}
\caption{{\bf Yielding due to persistent active forces.}  
(A) Scaling collapse of the response $R$ of a particle packing to an applied active force $f$, similar to what is seen in sheared dense particle simulations~\cite{liao2018criticality}.  
(B,C) Magnitude of inter-particle forces seen in active particle packings in the limit that the persistence time gets very large, at two different values of the magnitude of active forces~\cite{mandal2020extreme}. 
(D-F) Comparison of avalanches in simulations of particle packings subject to simple shear (D) or infinitely persistent random forces (E), from Ref~\cite{morse2020direct}. 
(F) As predicted by mean-field theories for glass dynamics, the avalanche statistics can be collapsed across sheared and active systems by scaling the data by $\kappa$, which depends on the correlation length of the input field (i.e. the size of the box for applied shear or the size of a particle for the active matter field).
}
\label{fig:active*}
\end{figure*}

In addition to finite strain rates and temperatures, many biological materials are also subject to some sort of active forces that are internally generated. In most cases, these forces are ultimately generated by molecular motors or other proteins that change their configuration to release stored energy. At the scale of fiber networks or cells, such forces self-organize to drive fluctuations in tension along edges of the network, or drive the self-propulsion of cells. Unlike thermal fluctuations, the active fluctuations can be characterized both by their magnitude and by some persistence time, the latter having no analog in equilibrium systems. Importantly, these types of persistent internal forces occur in the confluent tissues discussed above, but also in non-confluent cell assemblies that are best described by metric interactions, and in artificial and biomimetic systems such as Janus colloids, magnetically driven beads, and \emph{in vitro} mixtures of fibers and motor proteins.

These materials have generically been termed active matter, and many of their interesting properties at low and intermediate densities are discussed in previous reviews~\cite{marchetti2013hydrodynamics, cates2015motility}. At higher densities, it has been shown that the glass transition in active matter can be different from that driven by thermal fluctuations~\cite{berthier2014nonequilibrium}. Here, we review results on the yielding behavior and rheology of active matter at high densities, in the limit of slow driving and small fluctuations.

There are two different ways to conceptualize an experiment to study the yielding of active matter. The first is to perform a standard rheology experiment, such as shearing the boundaries (macroscopic rheology) or driving a tracer particle through a bath of active particles (micro-rheology). The micro-rheology approach was recently taken to investigate active monodisperse disks, and the authors find power-law-distributed velocity time series that are associated with intermittent, avalanche-like behavior~\cite{reichhardt2015active}.

A second approach is to note that in the limit in which the active particles are highly persistent (i.e. the rotational fluctuations are small), the material becomes self-shearing~\cite{briand2018spontaneously}, in the sense that the active forces that act at the local scale now behave as some sort of mechanical forcing, in analogy with the mechanical deformations usually driving the system at large scale. In this approach the macroscopic rheology of the system is not probed at all, but the competition between the large density of the particle system with the local driving force applied to each particle can drive the relaxation of the system in a way that is reminiscent of the yielding transition.  

This second approach has been used by several groups. In the limit of zero rotational noise, the active self-propulsion forces effectively become a quenched random field of forces applied on each particle. In this view, the average amplitude of those random forces plays a role similar to the shear stress in a traditional rheological experiment. Under such field, Liao and Xu~\cite{liao2018criticality} find a scaling collapse of the response function on either side of a critical jamming density; in the limit of slow driving they show that above the critical density the system behaves as a yield stress solid, and below it behaves as a fluid with a finite viscosity. This scaling collapse is shown in Fig.~\ref{fig:active*}A, and is very similar to rheological observations near the jamming transition~\cite{olsson2007critical}. Mandal {\it et al.} studied the response of active matter to a wide range of values for the particle persistence. In the limit of infinite persistence at high densities they similarly find a yielding solid state~\cite{mandal2020extreme}. Examples of the force chains that exist in the solid between rearrangement events are shown in Figs.~\ref{fig:active*}B,C for two different values of the magnitude of the active force.

Recently, Morse and coworkers investigated whether this analogy could be made more precise by explicitly studying avalanche statistics in the pre-yielding regime in response to both shear, Fig.~\ref{fig:active*}D, and quenched random forces, Fig.~\ref{fig:active*}E, in the quasi-static limit of infinitely slow driving~\cite{morse2020direct}. One motivation for this approach is the possibility to derive exact dynamical equations for infinite-dimensional particles, where the functional form of the equations suggests that in mean-field, forces applied randomly to each particle behave similarly to shear forces applied at the boundary~\cite{agoritsas2019out}. In recent work~\cite{morse2020direct, agoritsas2021mean}, the authors demonstrate explicitly that in infinite dimensions, the avalanche statistics (including the sizes of events and the local shear modulus associated with elastic branches) can be collapsed by a single scaling factor related to the correlation length of the applied field, as shown in Fig.~\ref{fig:active*}F. In sheared systems, the correlation length of the field is the size of the box, while in the quenched random field it can be as small as the distance between two particles. 

This scaling collapse was initially identified in infinite dimensions, but it also holds in 2D simulations of jammed soft particles. In other words, a random quenched force field applied to every particle generates a response that is identical to applied shear up to a simple scaling factor. As the correlation of the random field increases toward the size of the box, the scaling factor approaches unity. This indicates that in the pre-yielding regime, the infinite-persistence limit of active matter can be described by the same tools as sheared systems.

Interestingly, there are hints that this equivalence breaks down at the yielding transition itself. Under shear, highly stable computer glasses generated using swap Monte Carlo exhibit brittle failure with localized shear bands and a large stress drop (Sec.~\ref{sec:yielding} above), but under applied random forces those same materials fail more gradually, with no obvious shear bands~\cite{morse2020direct}. Future work should focus on understanding whether there is some less-obvious strain localization in these systems, whether symmetries in the applied field are necessary for brittle failure, or whether there are strong finite-size effects preventing their numerical observation. Additional open questions include understanding how adding small amounts of rotational noise, or finite driving rates to active matter systems perturbs them away from this exact analogy with sheared systems.

When active forces are highly persistent but a small amount of rotational noise is introduced, the system can flow even when the magnitude of the applied active forces is below the yielding threshold discussed before, and this driven dynamics again displays qualitative similarities with sheared materials~\cite{keta2022intermittent}. In this limit, the system travels through a sequence of mechanical equilibria where particle interaction and active forces compensate each other over very long periods of time. Due to the small rotational noise, however, there comes a moment when active forces have significantly evolved, and the system may then suddenly transition to a new equilibrium between interaction and active forces via a large-scale avalanche. In this limit, the dynamics driven by highly persistent forces reaches a dynamic steady state characterised by intermittent relaxation events due to avalanches separated by elastic response. Numerical results~\cite{keta2022intermittent} indicate that despite qualitative similarities, the statistics of avalanches or the nature of plastic and elastic responses may differ quantitatively from boundary driven shear flows, which raises interesting theoretical challenges for future work.   

\section{Conclusion}

Our primary emphasis has been on establishing a unified theoretical framework that interconnects the varied rheological behaviors of amorphous systems, with glass stability serving as an organizing principle. We have also explored new research areas where rheology is playing an increasingly important role such as biological tissues and active matter. Future research is needed to better characterize the common and universal rheological properties of these systems but also delving into the unique attributes that distinguish each one of them.

\acknowledgments

We thank all members and affiliates of the Simons collaboration, and the whole community, for many discussions over the years. This work was supported by a grant from the Simons Foundation (\#454933, Ludovic Berthier, \#454935 Giulio Biroli, \#454947 Lisa Manning, \#454955 Francesco Zamponi ).

\bibliography{main.bib}

\end{document}